\pdfoutput=1
\documentclass[twocolumn,superscriptaddress,showpacs,prl]{revtex4}
\usepackage{mathrsfs}
\usepackage{amssymb, amsbsy, amsmath, latexsym, dsfont, array, layout, graphicx,mathrsfs,color}

\newcommand{\one}{\mathds{1}}
\newcommand{\ket}[1]{\left|{#1}\right\rangle}
\newcommand{\bra}[1]{\left\langle{#1}\right|}

\begin{document}

\title{Realization of single-qubit positive operator-valued measurement
	via a one-dimensional photonic quantum walk}
\author{Zhihao Bian}
\affiliation{Department of Physics, Southeast University, Nanjing
211189, China}
\author{Jian Li}
\affiliation{Department of Physics, Southeast University, Nanjing
211189, China}
\author{Hao Qin}
\affiliation{Department of Physics, Southeast University, Nanjing
211189, China}
\author{Xiang Zhan}
\affiliation{Department of Physics, Southeast University, Nanjing
211189, China}
\author{Rong Zhang}
\affiliation{Department of Physics, Southeast University, Nanjing
211189, China}
\author{Barry C. Sanders}
\affiliation{%
    Hefei National Laboratory for Physical Sciences at Microscale and Department of Modern Physics,
    University of Science and Technology of China, Hefei, Anhui 230026, China
    }
\affiliation{%
    Shanghai Branch,
    CAS Center for Excellence and Synergetic Innovation Center
        in Quantum Information and Quantum Physics,
    University of Science and Technology of China, Shanghai 201315, China
    }
\affiliation{%
    Institute for Quantum Science and Technology, University of Calgary, Alberta, Canada T2N 1N4
    }
\affiliation{%
    Program in Quantum Information Science,
    Canadian Institute for Advanced Research,
    Toronto, Ontario M5G 1Z8, Canada
    }
\author{Peng Xue\footnote{gnep.eux@gmail.com}
} \affiliation{Department of Physics, Southeast University, Nanjing
211189, China}\affiliation{State Key Laboratory of Precision
Spectroscopy, East China Normal University, Shanghai 200062, China}
\affiliation{Beijing Institute of Mechanical and Electrical Space, Beijing 100094, China}
\begin{abstract}
We perform generalized measurements of a qubit by realizing the qubit as a coin in a photonic quantum walk and subjecting the walker to projective measurements. Our experimental technique can be used to realize photonically any rank-1 single-qubit positive operator-valued measure via constructing an appropriate interferometric quantum-walk network
and then projectively measuring the walker's position at the final step.
\end{abstract}

\maketitle

Quantum walks (QWs) exhibit distinct features compared to classical
random walks with applications to
quantum algorithms~\cite{Kem03a,Ven12}.
The discrete-time QW
is a process in which the evolution of a quantum particle on a
lattice depends on a state of a coin,
typically a two-level system, or qubit.
Controlling the coin degree of freedom
indirectly controls the walker,
and, through this indirect control,
the walker's state can be measured to infer the coin state.
Rigorously speaking,
walker-coin entanglement and projective measurement of the walker yields
a positive operator-valued measure (POVM) on a single qubit~\cite{KW13}.
Furthermore any rank-1 and rank-2 single-qubit POVM
can be generated by a judiciously engineered QW. Here we demonstrate experimentally the capability
of performing such generalized measurements of a qubit by
realizing the walker in the path degree of freedom of a photon and the coin state as polarization
and performing optical interferometry with path-based photodetector to perform a POVM on the photon's polarization state.

Realizing a POVM is important as a POVM is needed
for generalized acquisition of information thereby associated
with a multitude of quantum information tasks
such as quantum state estimation and tomography~\cite{AJK05},
quantum cloning~\cite{SIGA05}, entanglement distillation~\cite{BBP+96}
and generalized quantum cryptography protocols~\cite{GRTZ02}.
Single-qubit POVMs have been performed experimentally~\cite{HMG+96,CCBR01,MSB04,MCWB06}.
POVMs' wide applications include
unambiguous state discrimination~\cite{Iva87,Die88,Per88,DSK05} and quantum
state tomography in terms of symmetric informationally
complete (SIC) POVMs~\cite{RBSC04,SG10}.

Our goal is to realize experimentally a single-qubit POVM
and to discriminate between non-orthogonal initial coin states
via executing a properly engineered QW
whose projective walker measurement is sometimes inconclusive~\cite{KW13}.
To achieve a site-specific POVM,
we control the internal degree of freedom of the measured two-level coin.
Here we report our successful experimental realization of POVMs,
including unambiguous state discrimination of two equally probable single-qubit states and a single-qubit SIC-POVM,
via a one-dimensional photonic QW.

We focus on rank-1 POVMs, as higher-rank POVMs can be constructed as a convex combination of rank-1 elements~\cite{KW13}. Our experimental technique can be used to realize photonically any rank-1 single-qubit POVM via constructing an interferometric QW and projectively measuring the walker's position at the final step.
We characterize experimental performance by the 1-norm distance~\cite{BFL+10}
between the walker distribution obtained experimentally~$P^\text{exp}(x)$ vs theoretically~$P^\text{th}(x)$
over integer-valued position~$x$.
This distance is
\begin{equation}
	d=\frac{1}{2}\sum_x\left|P^\text{exp}(x)-P^\text{th}(x)\right|,
\end{equation}
and a small distance indicates a successful experimental realization.

A standard model of a one-dimensional (1D) discrete-time QW consists of a walker
carrying a coin that is flipped before each step. In the coin-state basis $\{\ket{0},\ket{1}\}$,
the site-dependent coin rotation for the $n^\text{th}$ step $C_{x,n}\in SU(2)$ is applied to the coin when the walker in the position~$x$,
followed by a conditional position shift due to the outcome of the coin
flipping for each step
$T=\sum_{x}\ket{x+1}\bra{x}\otimes\ket{0}\bra{0}+\ket{x-1}\bra{x}\otimes\ket{1}\bra{1}$.
The unitary operation for the $n^\text{th}$ step is
$U_n=T\sum_x\ket{x}\bra{x}\otimes C_{x,n}$.

We commence with the simplest non-trivial case,
namely a three-step QW approach to implement an unambiguous state discrimination of two single-qubit states. Two nonorthogonal pure states can always be encoded as
\begin{equation}
\label{eq:phipm}
	\ket{\phi^\pm}=\cos \frac{\phi}{2}\ket{0}\pm\sin\frac{\phi}{2}\ket{1}.
\end{equation}
Our objective is to discriminate these two states with equal prior probability
for three outcomes:
conclusively measuring one or the other state~(\ref{eq:phipm})
or obtaining an inconclusive measurement result.
We prepare an initial coin state in either of the two states~(\ref{eq:phipm}).
For a properly engineered QW procedure,
the walker with the different initial coin states arrives at different position distributions.
By projective measurement onto the walker's position,
initial coin states can be discriminated.

\begin{figure}
\includegraphics[width=\columnwidth]{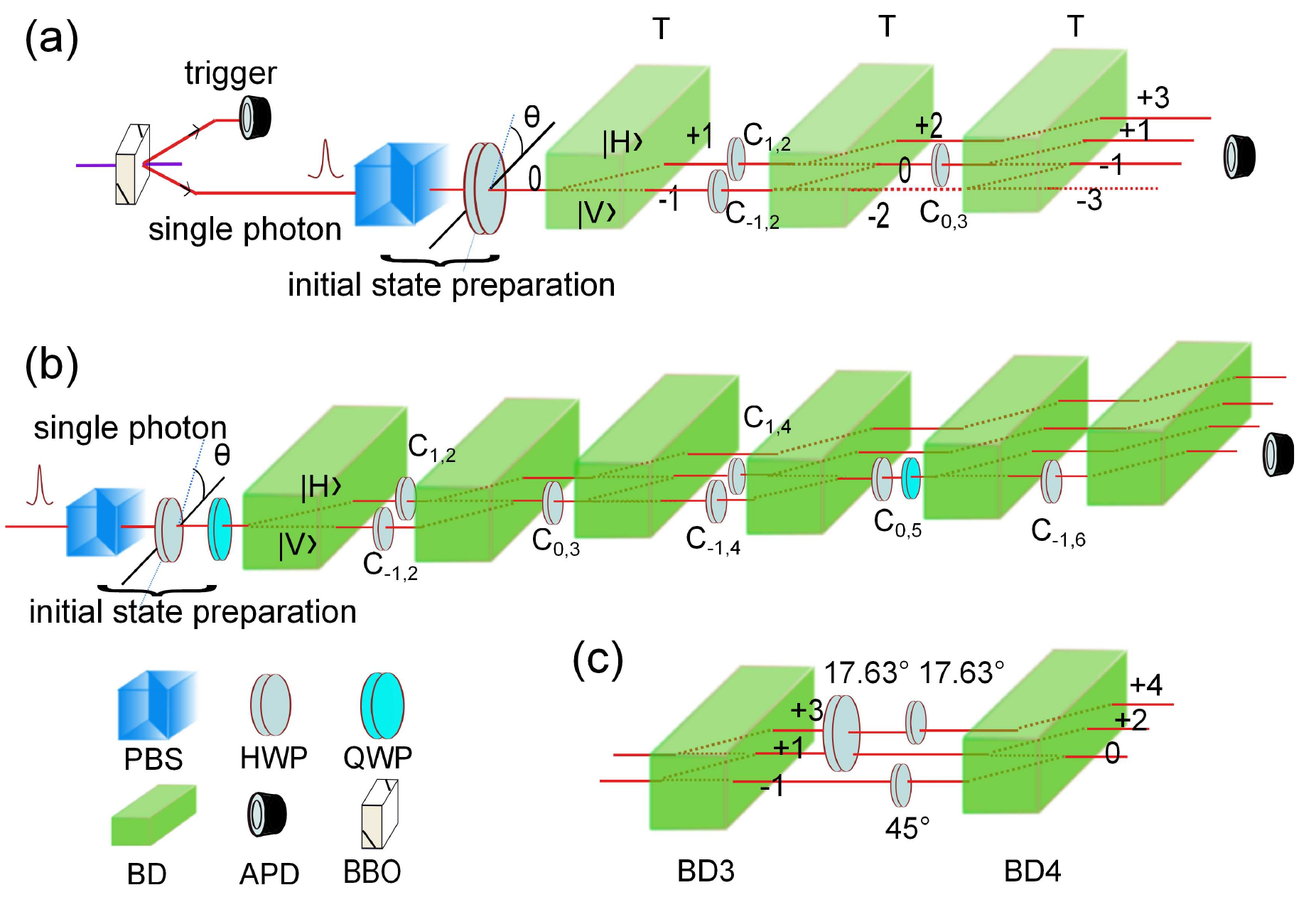}
\caption{%
	(Color online.) Experimental schematic.
	(a)~Detailed sketch of
the setup for realization of unambiguous state discrimination of two
equally probable single-qubit states via a three-step QW.
Single-photons are created via SPDC in a BBO crystal. One photon in the pair is detected to herald the other photon, which is injected into the optical network.
	(b)~Setup for realization of a qubit SIC-POVM via a six-step QW.
The initial coin states for realization of SIC-POVM~$\ket{\psi^{1,2}}$
are prepared by a HWP, whereas~$\ket{\psi^{3,4}}$ are prepared by a QWP with certain setting angles.
	Site-dependent coin flipping is realized by HWPs
	(and QWPs) with different setting angles placed in different
optical modes. (c)~Detailed interferometric setup formed by the third and fourth BDs,
which are used in the experimental realization of a qubit SIC-POVM.%
	}
\label{setup}
\end{figure}

For our realization,
the coin is initially prepared in~$\ket{\phi^\pm}$ and the walker starts from
the origin $\ket{x=0}$. The site-dependent coin rotations
for the first three steps are
\begin{align}
	C_{-1,2}=&\begin{pmatrix}
                 \sqrt{1-\tan^2{\frac{\phi}{2}}} & \tan{\frac{\phi}{2}} \\
                 \tan{\frac{\phi}{2}} & -\sqrt{1-\tan^2{\frac{\phi}{2}}} \\
               \end{pmatrix}, \nonumber\\
	C_{1,2}=&\sigma_x=\begin{pmatrix}
                 0 & 1 \\
                 1 & 0 \\
              \end{pmatrix},
	C_{0,3}=\frac{1}{\sqrt{2}}\begin{pmatrix}
                 1 & 1 \\
                 1 & -1 \\
              \end{pmatrix}
\end{align}
and~$\one$ elsewhere. Each step the site-dependent coin rotations are followed by a conditional position shift $T$.
Then the initial walker-coin states $\ket{\varphi^\pm_0}=\ket{0}\ket{\phi^\pm}$ evolve into
\begin{align}
&\ket{\varphi^+_3}=\sqrt{\cos
\phi}\ket{3}\ket{0}+\sqrt{2}\sin\frac{\phi}{2} \ket{1}\ket{0},\nonumber\\
&\ket{\varphi^-_3}=\sqrt{\cos
\phi}\ket{3}\ket{0}-\sqrt{2}\sin\frac{\phi}{2} \ket{-1}\ket{1}
\end{align}
respectively.
The walker is projectively measured in the position basis.
If the position measurement outcome is $x=1$ ($x=-1$),
then we ascertain that the initial coin state was~$\ket{\phi^+}$ ($\ket{\phi^-}$).
If the walker is instead measured in $x=3$,
we do not know the initial coin state;
that is, $x=3$ corresponds to an inconclusive result with probability
$\eta_\text{err}=|\bra{\phi_+}\phi_-\rangle|=\cos\phi$.
The probability of the inclusive result depends on the similarity of the two states,
which agrees with the Ivanovic-Dieks-Peres bound
obtained from the optimum strategy of this kind for unambiguous state
discrimination~\cite{Iva87,Die88,Per88}.
For the extreme case of $\phi=\pi/2$,
the measurement is
projective and the probability~$\eta_\text{err}$ is $0$.

Now we establish the universality of the QW
procedure for generation of an arbitrary rank-1
POVM via an example of a properly engineered six-step QW generating a single-qubit SIC-POVM. For example we choose
\begin{align}
\label{eq:qubitsicpovm}
&\ket{\xi^1}=\ket{0},
\ket{\xi^2}= \frac{1}{\sqrt{3}}\left(\ket{0}+\sqrt{2}\ket{1}\right),\\
&\ket{\xi^3}=\frac{1}{\sqrt{3}}\left(\ket{0}+\lambda\sqrt{2}\ket{1}\right),
\ket{\xi^4}=\frac{1}{\sqrt{3}}\left(\ket{0}+\lambda^*\sqrt{2}\ket{1}\right)\nonumber
\end{align}
satisfying $|\langle \xi^i|\xi^j\rangle|=3^{-1/2}$ for $i\neq j$
and $\frac{1}{2}\sum_i^4\ket{\xi^i}\bra{\xi^i}=\one$
for $\lambda=\text{e}^{\text{i}2\pi/3}$.
Now we construct four states orthogonal to the above states~(\ref{eq:qubitsicpovm})
and prepare the coin state in one of the four states initially.
Thus, the QW procedure starts with four initial coin states
\begin{align}
&\ket{\psi^1}=\ket{1},
\ket{\psi^2}=\frac{1}{\sqrt{3}}\left(\sqrt{2}\ket{0}-\ket{1}\right),\nonumber\\
&\ket{\psi^3}=\frac{1}{\sqrt{3}}\left(\sqrt{2}\ket{0}-\lambda\ket{1}\right),\\
&\ket{\psi^4}=\frac{1}{\sqrt{3}}\left(\sqrt{2}\ket{0}-\lambda^*\ket{1}\right). \nonumber
\end{align}
The site-dependent coin rotations for the first six steps are
\begin{align}
	&C_{1,2}=\frac{1}{\sqrt{2}} \begin{pmatrix}
           1 & -1 \\
           -1 & -1 \\
        \end{pmatrix},
	C_{0,3}=\frac{1}{\sqrt{2}}\begin{pmatrix}
           -1 & 1 \\
           1 & 1 \\
        \end{pmatrix},
        		\nonumber\\
&C_{-1,2}=C_{-1,4}=C_{-1,6}=\sigma_x,\\
&C_{1,4}=\frac{1}{\sqrt{3}}\begin{pmatrix}
           \sqrt{2} & 1 \\
           1 & -\sqrt{2} \\
         \end{pmatrix},
       C_{0,5}=\frac{1}{\sqrt{2}}\begin{pmatrix}
           \text{e}^{-\text{i}\frac{\pi}{3}} & \text{e}^{\text{i}\frac{\pi}{6}} \\
           \text{e}^{\text{i}\frac{\pi}{3}} & \text{e}^{-\text{i}\frac{\pi}{6}} \\
                   \end{pmatrix},\nonumber
\end{align}
and $\one$ elsewhere.
The coin operators chosen here depend only on the states we aim to discriminate.
Following the six-step QW procedure including specific site-dependence coin rotations,
the initial states of the walker-coin system
$\ket{\varphi^i_0}=\ket{0}\ket{\psi^i}$
($i=1,2,3,4$)
evolve to
\begin{align}
	 &\ket{\varphi^1_6}=\frac{1}{\sqrt{3}}\left(-\ket{4}-\text{i}\ket{2}+\text{i}\ket{0}\right)\ket{0},\nonumber\\
	 &\ket{\varphi^2_6}=\frac{1}{\sqrt{3}}\left(\ket{6}-\text{e}^{-\text{i}\frac{\pi}{3}}\ket{2}-\text{e}^{\text{i}\frac{\pi}{3}}\ket{0}\right)\ket{0},\nonumber\\
	 &\ket{\varphi^3_6}=\frac{1}{\sqrt{3}}\left(\ket{6}-\text{e}^{-\text{i}\frac{\pi}{6}}\ket{4}-\ket{2}\right)\ket{0},\nonumber\\
	 &\ket{\varphi^4_6}=\frac{1}{\sqrt{3}}\left(\ket{6}-\text{e}^{\text{i}\frac{\pi}{6}}\ket{4}-\ket{0}\right)\ket{0}.
\end{align}
Evidently the final walker-coin states have differing support over position~$x$ so
by measuring the position of the walker,
the initial coin state can be determined with some degree of certainty.
Specifically,
$\ket{\varphi^{1,2,3,4}_6}$ cannot be found at $x=6$,
$x=4$, $x=0$ and $x=2$, respectively.
Thus, we realize all elements
$\ket{\xi^i}\bra{\xi^i}/2$ ($i=1,2,3,4$) of a qubit SIC-POVM
through a QW procedure.

\begin{figure}
\includegraphics[width=\columnwidth]{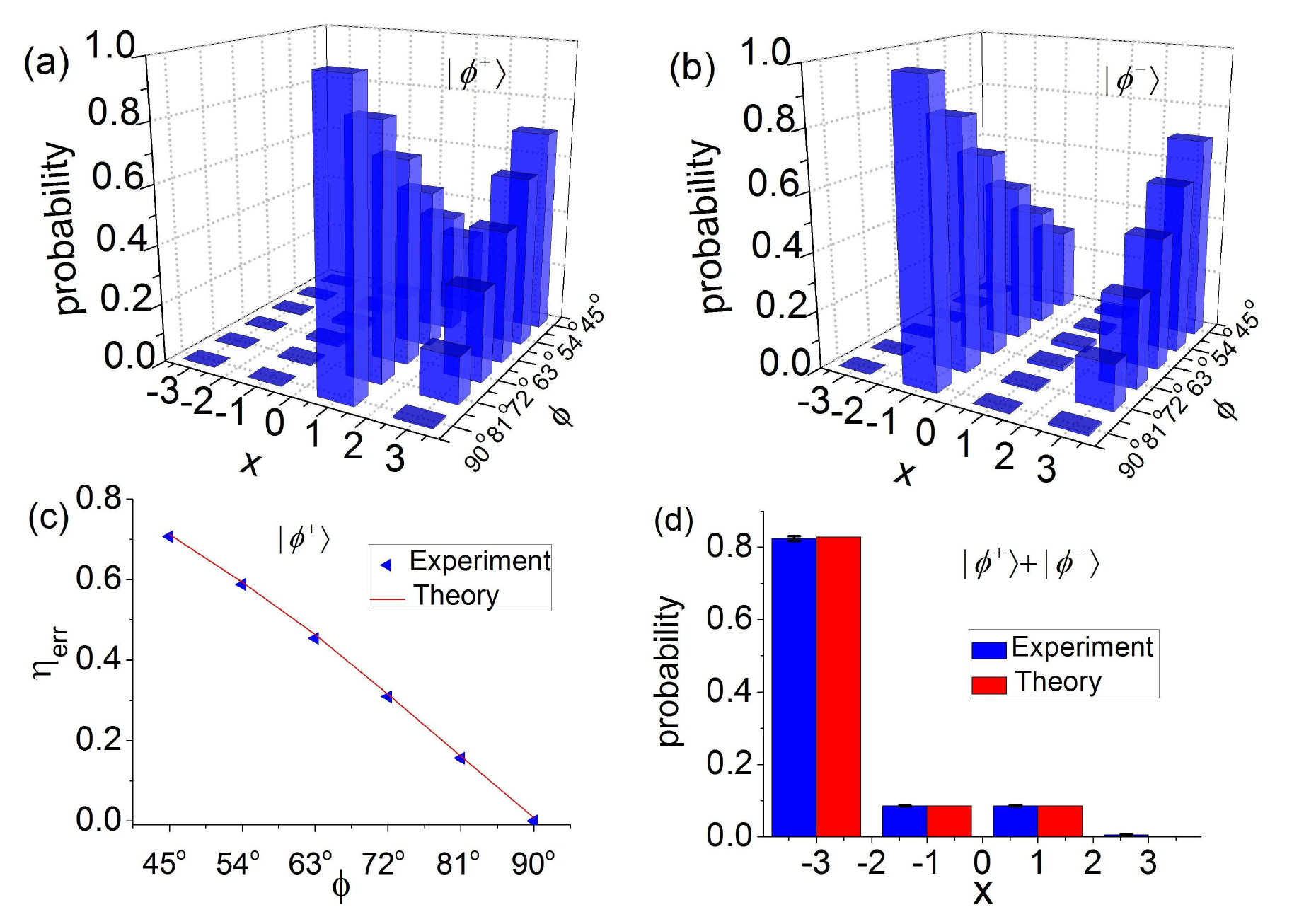}
\caption{%
	(Color online.)
	Experimental data for unambiguous state discrimination via a photonic QW.
	Measured position distributions for the three-step QW
	with site-dependent coin and (a)~initial coin state~$\ket{\phi^+}$ and (b)~$\ket{\phi^-}$;
	various coefficients~$\phi$ of~$\ket{\phi^\pm}$ for unambiguous
state discrimination.
	(c)~Measured probability~$\eta_\text{err}$ for inconclusive results
	vs parameters~$\phi$,
	which are related to the state to be discriminated;
	compared to theoretical predictions.
	Error bars are smaller than portrayed by the symbols.
	(d)~Position distribution for the three-step QW with initial coin state~$\ket{H}$,
	which is an equally-weighted superposition of~$\ket{\phi^\pm}$ with $\phi=45^\circ$.
The blue and red bars show the experimental data and theoretical
predictions, respectively. Error bars indicate the statistical uncertainty.}\label{discrimination}
\end{figure}

\begin{figure}
\includegraphics[width=\columnwidth]{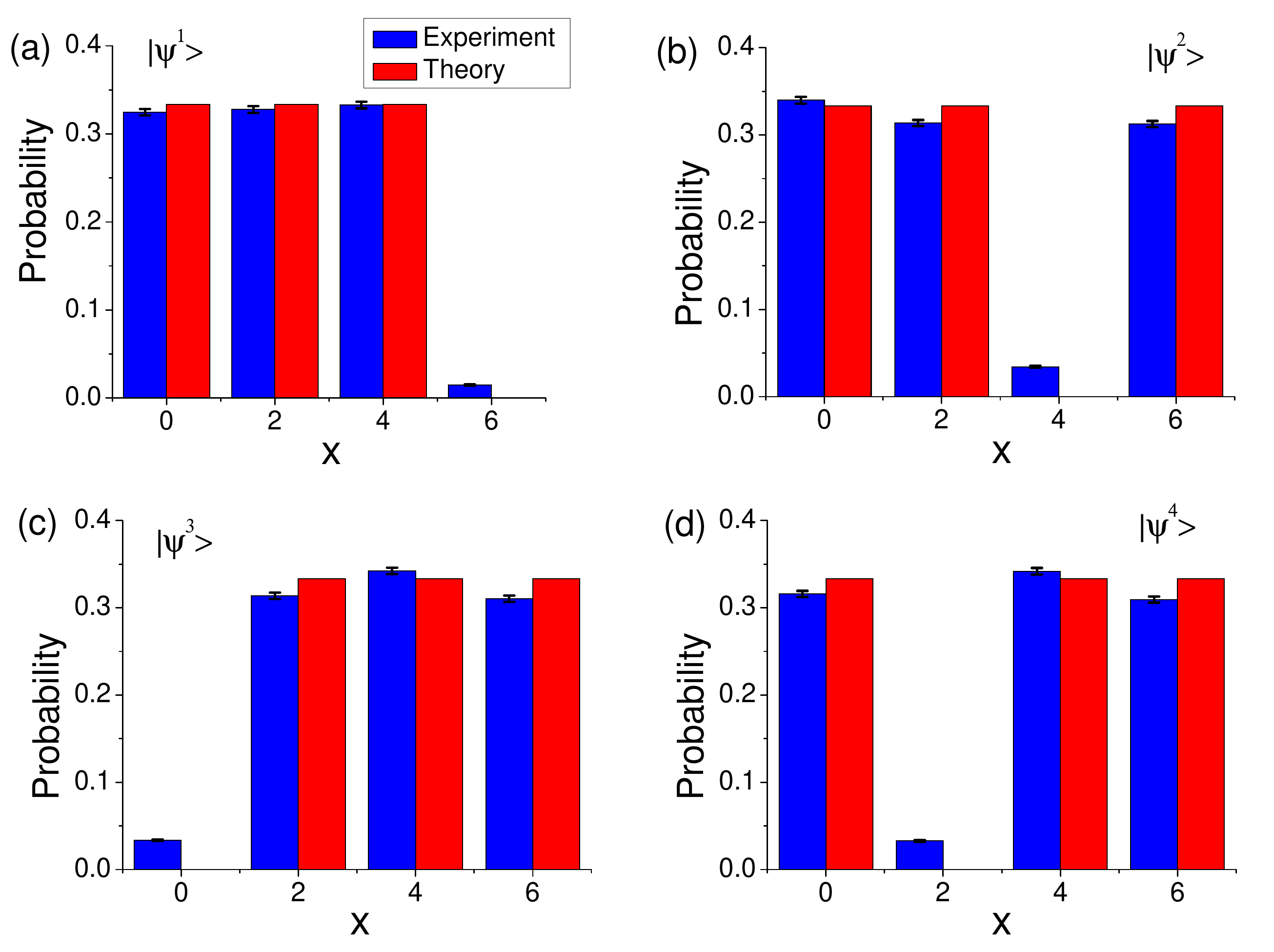}
\caption{(Color online.) Experimental data of a qubit SIC-POVM via a photonic QW.
Measured probability distributions of the six-step QW with the site-dependent coin rotations and four different initial
coin states $\ket{\psi^i}$ with $i=1,2,3,4$ in (a)-(d) respectively.}\label{POVM}
\end{figure}

The realization of unambiguous state discrimination of two equally
probable single-qubit states via a three-step QW is shown in Fig.~\ref{setup}(a). The coin qubit is encoded in the horizontal $\ket{H}=\ket{0}$ and vertical polarization $\ket{V}=\ket{1}$ of photons. The walker's positions are represented by longitudinal spatial
modes. The polarization degenerate photon pairs are generated via type-I
spontaneous parametric down-conversion (SPDC) in $0.5$mm-thick
nonlinear-$\beta$-barium-borate (BBO) crystal, which is pumped by a CW diode laser with $90$mW of power. For 1D QWs, triggering on one photon prepares the other beam at wavelength $801.6$nm into a single-photon state.

The initial coin state can be prepared by the half-wave plate (HWP) or quarter-wave plate (QWP) right after the polarizing beam splitter (PBS) shown in Fig.~\ref{setup}. After passing through the PBS and wave plate (WP) the down-converted photons are steered into the optical modes of the linear-optical network formed by
a series of birefringent calcite beam displacers (BDs) and WPs.
The site-dependent coin rotations $C_{x,n}$ for the $n^\text{th}$ step can be realized by HWP and QWP with specific setting angles placed in mode~$x$.

The conditional position shift is implemented by a BD with length $28$mm
and clear aperture $10$mm$\times 10$mm. The optical axis of each BD
is cut so that vertically polarized photons are directly transmitted
and horizontal photons move up a $2.7$mm lateral displacement into
a neighboring mode, which interfere with the vertical photons in the
same mode. Certain pairs of BDs form an interferometer, which are placed
in sequence and need to have their optical axes mutually aligned.

We attain
interference visibility of~0.992 for each step. Output photons are detected using avalanche photo-diodes
(APDs, $7$ns time window) with dark-count rate of less than 100s$^{-1}$
whose coincidence signals,
monitored using commercially available counting logic,
are used to post-select two single-photon events.
The walker position probabilities are obtained by
normalizing the coincidence counts on each mode with respect to the total count for each
respective step.

For site-dependent coin flipping, the optical delay usually needs
to be considered.
Fortunately, in our experiment on unambiguous state discrimination,
only the first and second BDs form an interferometer.
For the third step of the QW, the photons in mode $x=2$,
which are all in~$\ket{H}$, move up to mode $x=3$ after the last BD
and thus do not interfere with other photons. Thus, no optical compensate is needed and the difficulty of the realization of the experiment is decreased.

The measured probability distributions of a three-step QW for unambiguous state discrimination are shown in Fig.~\ref{discrimination}. We choose a different coefficient~$\phi$ and
prepare the initial coin state to the corresponding state
$\ket{\phi^\pm}$. For either
of the two states, the photons undergoing the QW
network are measured at the mode $x=3$ for inclusive result and
$x=\pm1$ for conclusive results. Two pronounced peaks for each
$\phi$ shown in the probability distribution in Figs.~\ref{discrimination}(a,b)
validate the demonstration of unambiguous state
discrimination. With~$\phi$ increasing from $45^\circ$ to $90^\circ$
the probability of inconclusive results~$\eta_\text{err}$ of the discrimination of
the state~$\ket{\phi^+}$ decreases from  $0.7139\pm0.0030$ to
$0.0070\pm0.0060$ (from $0.7125\pm0.0031$ to $0.0080\pm0.0071$ to
discriminate the state~$\ket{\phi^-}$) shown in
Fig.~\ref{discrimination}(c), agreeing with the Ivanovic-Dieks-Peres bound.

Taking a superposition of~$\ket{\phi^\pm}$ as an initial coin state
$a\ket{\phi^+}+b\ket{\phi^-}$ (non-normalized), with $a,b\in\mathbb{R}$,
the ratio of the probabilities for the two conclusive results is
$a^2/b^2$, which is also demonstrated in our experiment. In Fig.~\ref{discrimination}(d)
we show with $a=b$ the probabilities of $x=1$ and $x=-1$ are
measured approximately equal, i.e., $P(1)=0.0854\pm0.0015$ and
$P(-1)=0.0850\pm0.0015$.

We characterize the quality of the experimental QW by its 1-norm distance. The unambiguous state
discrimination is confirmed by direct measurement and found to be
consistent with the ideal theoretical values at the level of the small
average distance $d<0.02$ and the fidelity of the coin state
measured in the position ($x=\pm1$) $F>0.9911$.

The realization of a qubit SIC-POVM via a properly engineered six-step QW is shown in Fig.~\ref{setup}(b).
For site-dependent coin flipping, the challenge is placing the WP into a given optical mode
without influencing the photons in the other modes.
For example, in Fig.~\ref{setup}(c) for the fourth step, the polarizations of photons in modes $x=\pm1$ should be  rotated by a HWP with setting angle $\theta_H=17.63^\circ$ and $\theta_H=45^\circ$ respectively to realize the site-dependent coin rotations $C_{\pm1,4}$,
and the photons in those two modes interfere in mode $x=0$ at the fourth BD. Because of the small separations between the neighboring modes, it is difficult to inset a HWP in the middle mode $x=1$ and avoid the photons in the neighboring modes passing through it.

In our experiment, we place a HWP with $\theta_H=17.63^\circ$ in both modes $x=1$ and $x=3$ followed by a HWP with the same angle in mode $x=3$ and a HWP with $45^\circ$ in mode $x=-1$. Thus, the photons in modes $x=\pm1$ do not suffer an optical delay and interfere with each other with a high visibility. The polarizations of photons in mode $x=3$ are not changed after two HWPs with the same angle. The photons in mode $x=3$ do not interfere with those in the other modes, though there is optical delay between them.
Hence optical compensation is not required.

The measured probability distributions of six-step QW for a qubit SIC-POVM are shown in Fig.~\ref{POVM}, which agree well with the theoretical predictions. Using the experimental distribution of the QW with the initial coin state $\ket{\psi^1}$ as an example, after six steps the probability $P(6)$ is measured as $0.0149\pm0.0007$ and thus is very small compared to the probabilities of the photons being measured in the other modes, i.e., $P(0)=0.3246\pm0.0037,P(2)=0.3277\pm0.0038,P(4)=0.3327\pm0.0038$,
which ensures that one of the elements of a qubit SIC-POVM is realized successfully.
The small distance $d<0.043$
demonstrates strong agreement between theoretical and measured distribution after six steps.
Dominant sources of experimental errors are
inaccuracy of angles controlled by the WPs and imperfect non-unit visibility.

In summary we experimentally show that QWs are capable of
performing generalized measurements on a single qubit.
Our demonstration employs a novel photonic QW with
site-dependent coin rotation for realizing a generalized measurement~\cite{KW13}.
The key experimental advance to realize a QW-based generlalized measurement
is the application of site-dependent coin rotations to control the coin's internal dynamics
and thereby effect the evolution of the walker.
We have thus demonstrated a new and versatile approach to generalized qubit measurements via
photonic quantum walks.

\acknowledgements
We would like to thank P. Kurzy\'{n}ski, C.~F. Li and Y.~S. Zhang for
stimulating discussions.
This work has been supported by NSFC under
grants 11174052 and 11474049,
the Open Fund from SKLPS of ECNU,
the CAST Innovation fund,
NSERC, AITF, and the China 1000 Talents program.
\bibliography{../qw}

\section{Supplementary material}

\subsection{METHODS AND EXPERIMENTAL ANALYSIS FOR REALIZATION UNAMBIGUOUS STATE DISCRIMINATION}

For a photonic quantum-walk approach to realize an unambiguous state discrimination
of two equally probable single-qubit states, the initial coin states are
\begin{equation}
\ket{\phi^\pm}=\cos \frac{\phi}{2}\ket{H}\pm\sin\frac{\phi}{2}\ket{V},
\end{equation}
where $\ket{H}=(1,0)^\text{T}$ and $\ket{V}=(0,1)^\text{T}$ represent the horizontal and vertical polarization states of single photons. The initial state preparation can be realized by heralded single photons passing through a polarizing beam splitter (PBS) and a half-wave plate (HWP) with the angle between optical axis and horizontal direction setting to $\theta^{H}=\frac{\phi}{4}$. The single-qubit rotation realized by a HWP is
\begin{equation}
R_\text{HWP}(\theta^H)=\begin{pmatrix}\cos2\theta^H & \sin2\theta^H\\
\sin2\theta^H & -\cos2\theta^H\end{pmatrix},
\end{equation}
where $\theta^H$ is the angle between the optic axes of HWP and horizontal direction.

The site-dependent coin rotation $C_{x,n}$ can be realized by a HWP with a proper angle $\theta^H_{x,n}$ between the optical axis and horizontal direction. During the processing, only the coin rotation applied on photons in the
position $x=1$ for the second step
\begin{equation}
C_{1,2}(\theta^H_{1,2})=R_\text{HWP}(\frac{1}{2}\arccos{\sqrt{1-\tan^2\frac{\phi}{2}}})
\end{equation} depends
on the choice of the initial coin state which need to be discriminated, which decreases the difficulty of the experimental realization. The parameters of the setup and experimental data are shown in Table I.

\begin{center}
\begin{table}[htbp]
\begin{tabular}{l|l|l|l|l|l|l}
  \hline
$\phi$ & $\ket{\phi^{i=\pm}}$& $\theta^H_{-1,2}$ & $\theta^H_{1,2}$ & $\theta^H_{0,3}$ & $\eta_{err}$ & $d$ \\
  \hline\hline
$45^o$ & $\ket{\phi^+}$ &  $45^o$ & $12.23^o$ & $22.5^o$ & $0.7139(30)$ & $0.0171(46)$ \\ \hline
$54^o$ & $\ket{\phi^+}$  & $45^o$ & $15.32^o$ & $22.5^o$ & $0.5963(38)$ & $0.0184(47)$ \\ \hline
$63^o$ & $\ket{\phi^+}$ & $45^o$ & $18.9^o$ & $22.5^o$ & $0.4638(45)$ & $0.0192(47)$ \\ \hline
$72^o$ & $\ket{\phi^+}$ & $45^o$  & $23.3^o$ & $22.5^o$ & $0.3166(54)$ & $0.0127(46)$\\ \hline
$81^o$ & $\ket{\phi^+}$ & $45^o$ & $29.33^o$ & $22.5^o$ & $0.1616(62)$ & $0.0066(44)$ \\ \hline
$90^o$ & $\ket{\phi^+}$ & $45^o$ & $45^o$ & $22.5^o$ & $0.0060(70)$ & $0.0060(35)$  \\ \hline
$45^o$ & $\ket{\phi^-}$  & $45^o$ & $12.23^o$ & $22.5^o$ &$0.7125(31)$  & $0.0152(45)$ \\ \hline
$54^o$ & $\ket{\phi^-}$  & $45^o$ & $15.32^o$ & $22.5^o$ & $0.5934(39)$ & $0.0156(47)$ \\ \hline
$63^o$ & $\ket{\phi^-}$ & $45^o$ & $18.9^o$ & $22.5^o$ & $0.4635(45)$ & $0.0183(46)$ \\ \hline
$72^o$ & $\ket{\phi^-}$ & $45^o$  & $23.3^o$ & $22.5^o$ & $0.3146(55)$ & $0.0136(49)$\\ \hline
$81^o$ & $\ket{\phi^-}$  & $45^o$ & $29.33^o$ & $22.5^o$ & $0.1606(63)$ & $0.0071(43)$ \\ \hline
$90^o$ & $\ket{\phi^-}$ & $45^o$ & $45^o$ & $22.5^o$ & $0.0080(71)$ & $0.0080(36)$  \\ \hline
\end{tabular}
\caption{%
	The coefficients of the states to be discriminated (column 1), the
initial coin states (column 2), the corresponding parameters for the HWP
settings (columns 3-5), the experimental data for the probability of inconclusive results (column 6) and
1-norm distance from the theoretical predictions (column 7). Error bars
indicate the statistical uncertainty.}
\end{table}
\end{center}




\subsection{METHODS AND EXPERIMENTAL ANALYSIS FOR REALIZATION OF SIC-POVM}

\begin{figure*}
\includegraphics[width=0.5\columnwidth]{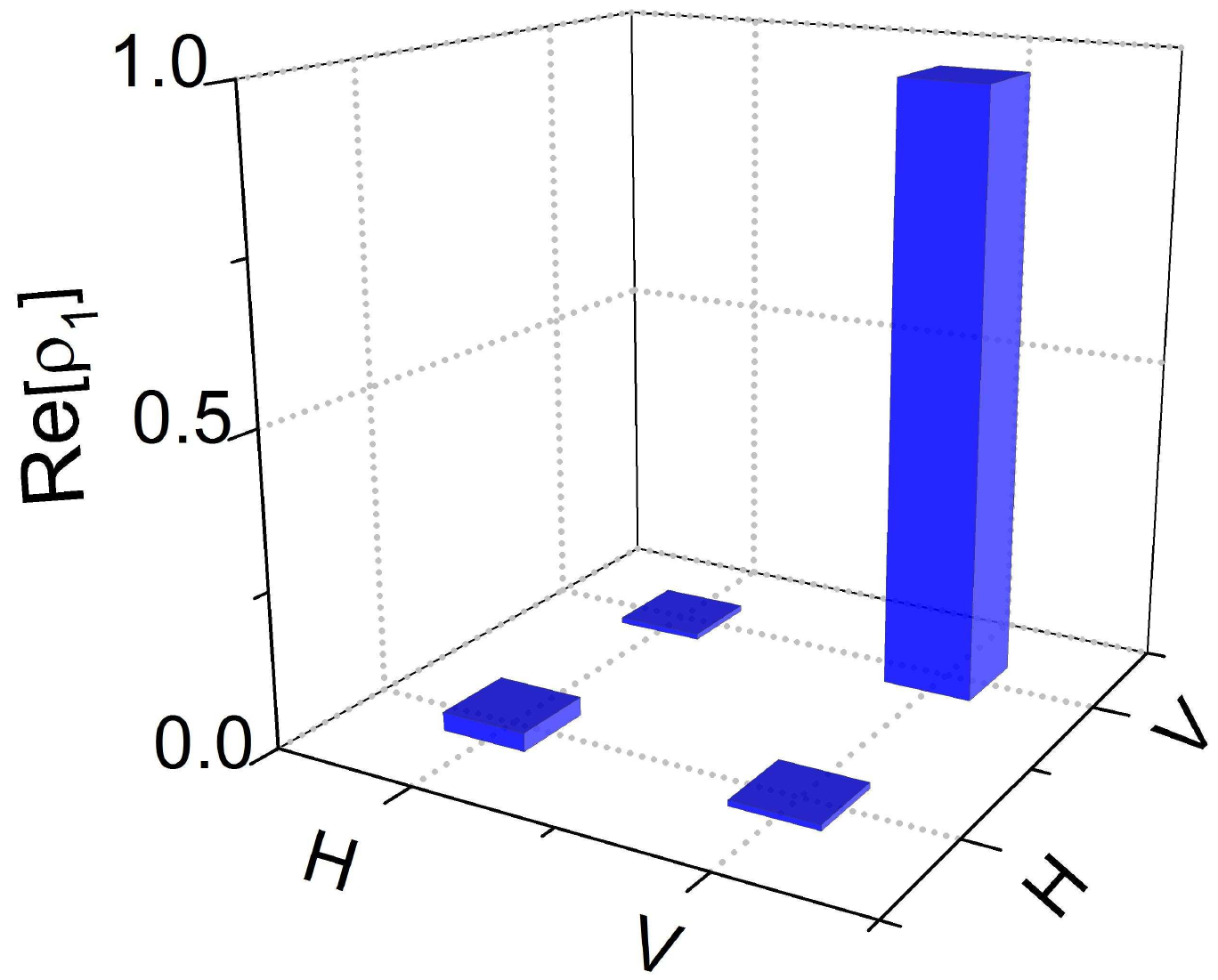}
\includegraphics[width=0.5\columnwidth]{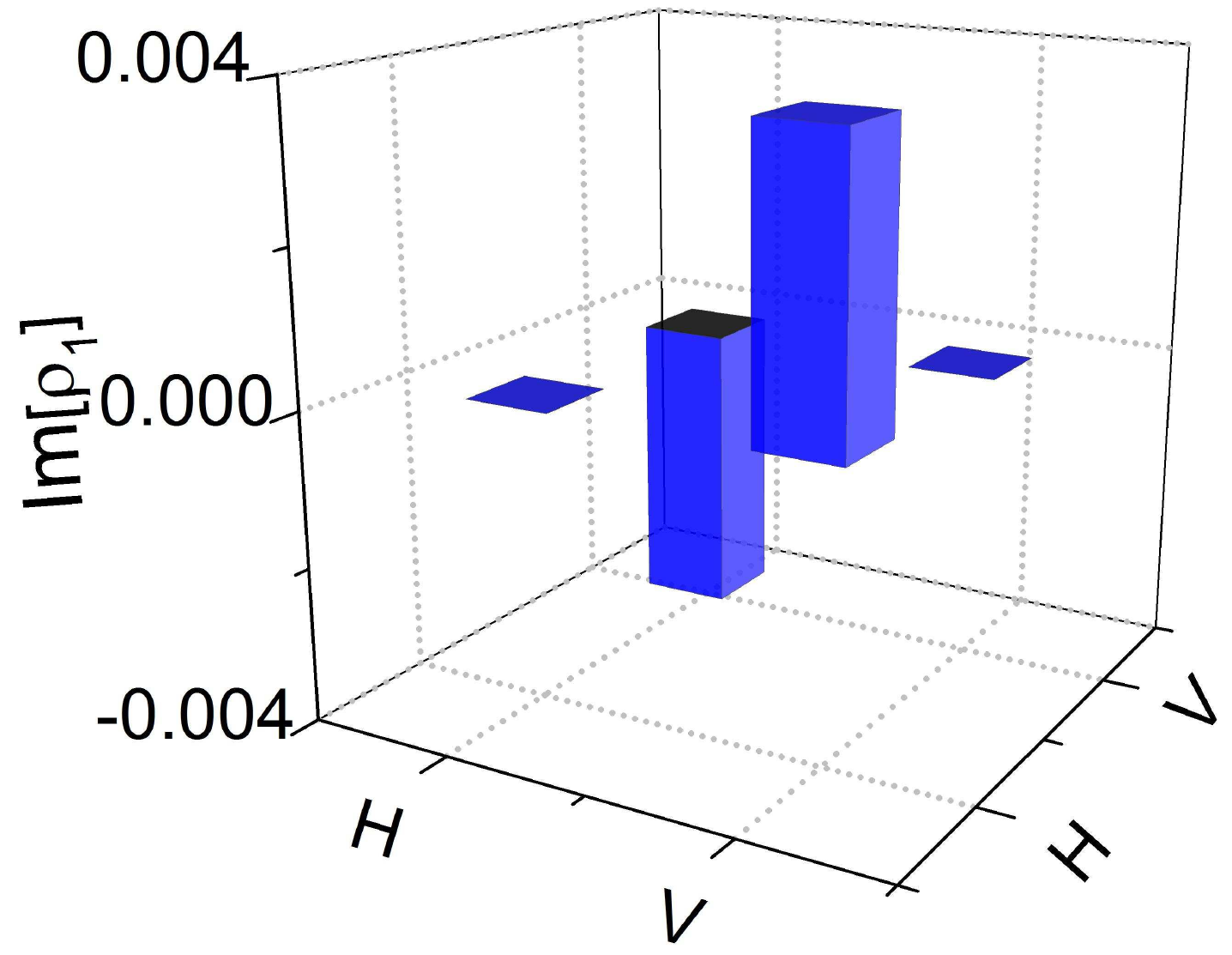}
\includegraphics[width=0.5\columnwidth]{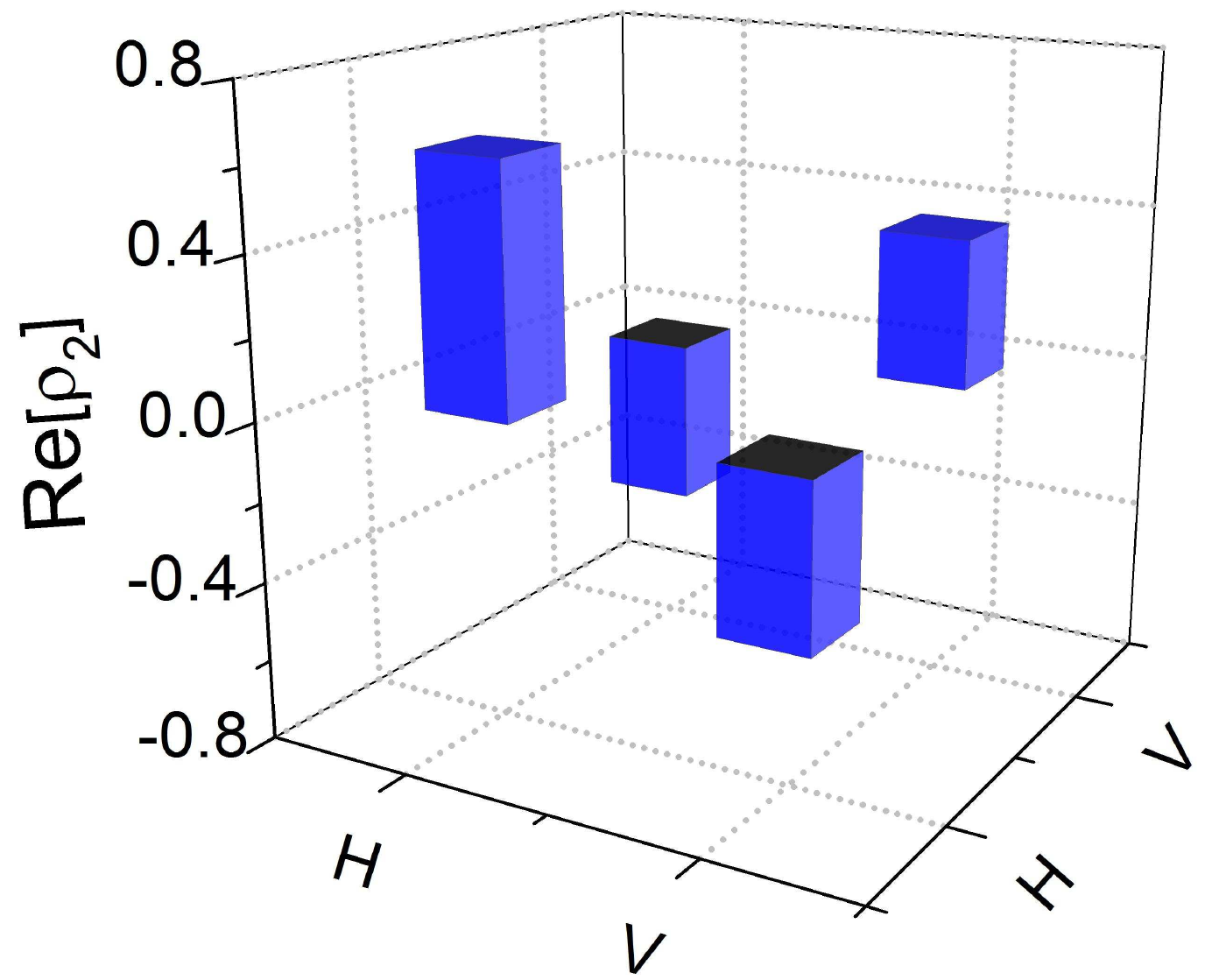}
\includegraphics[width=0.5\columnwidth]{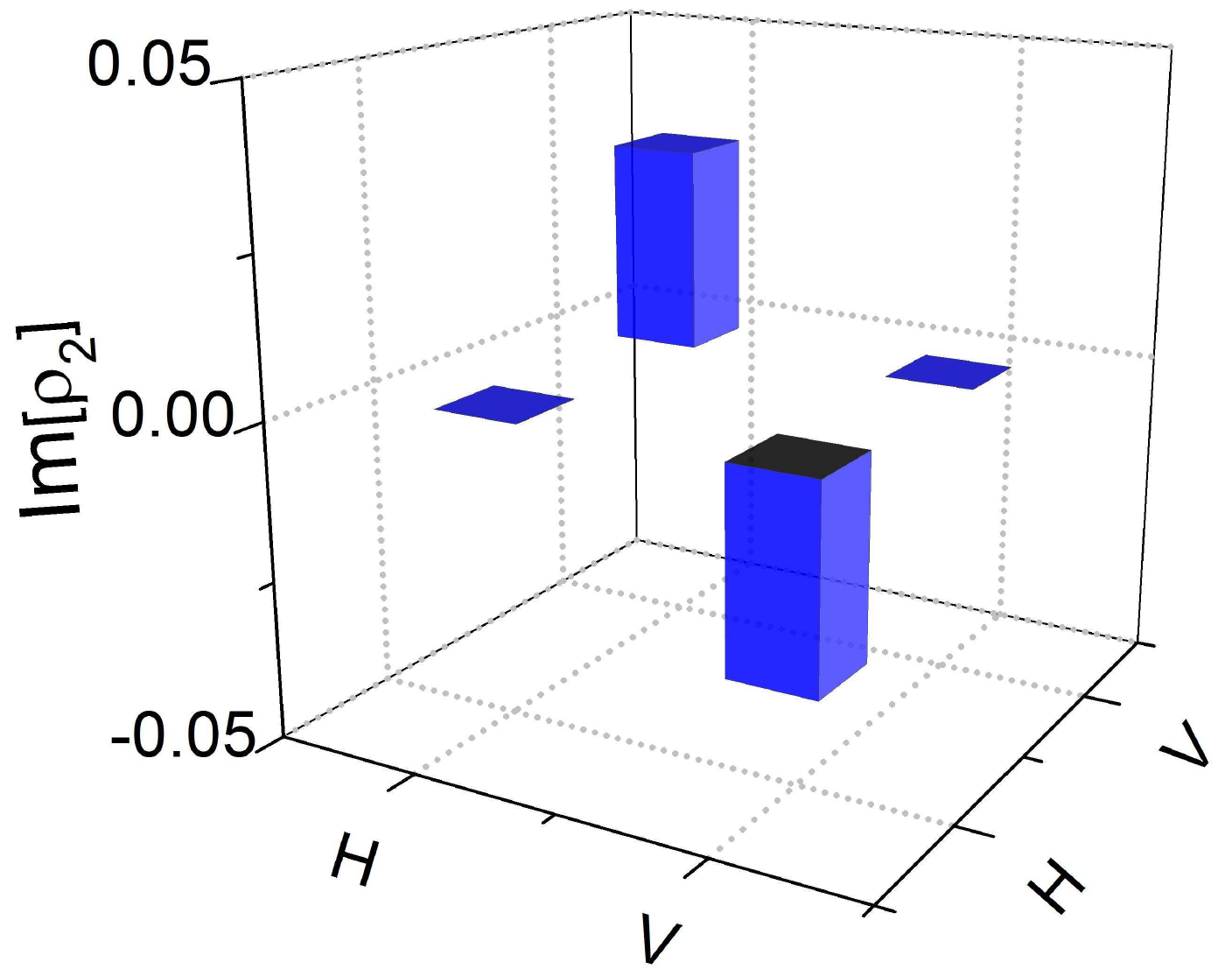}
\includegraphics[width=0.5\columnwidth]{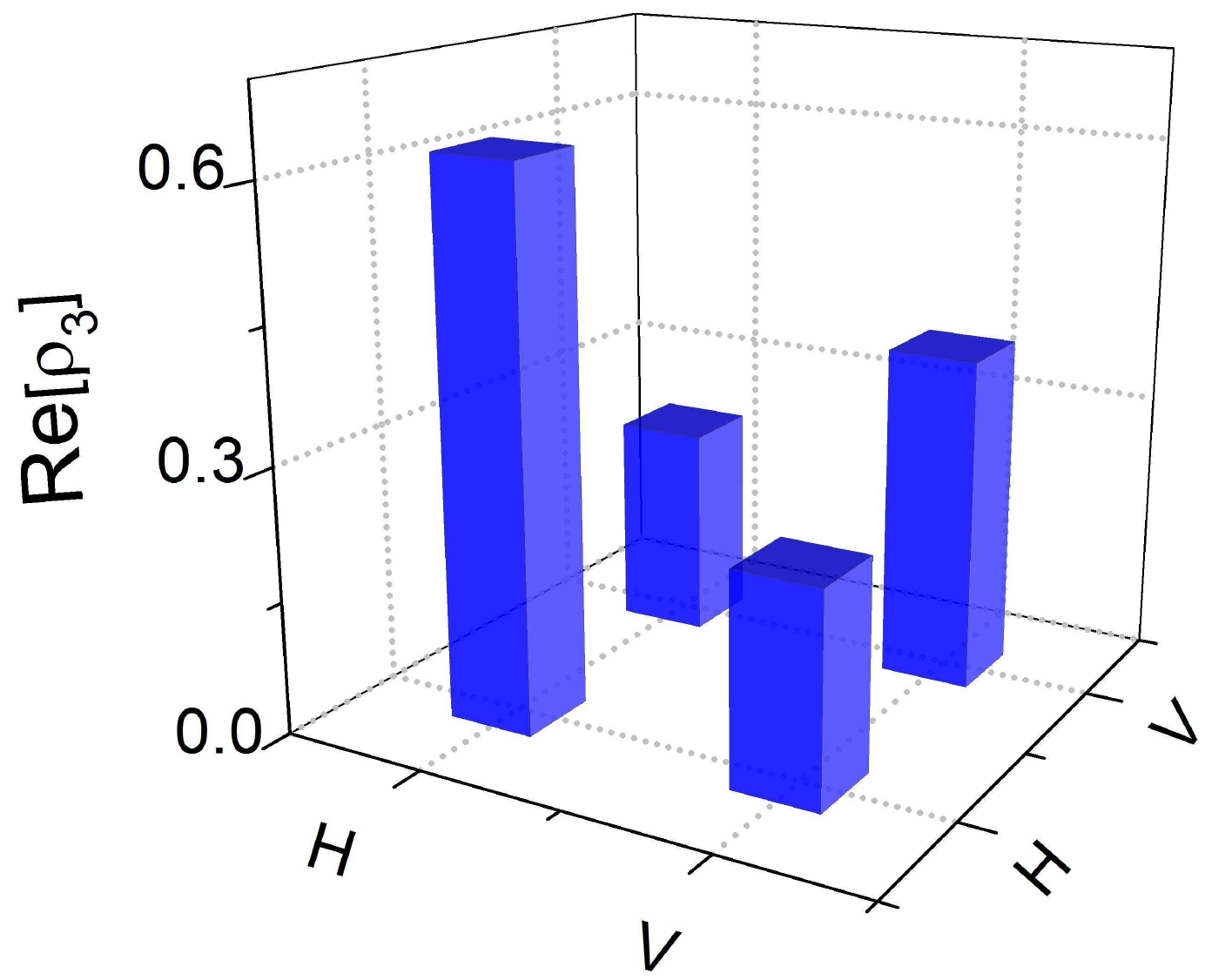}
\includegraphics[width=0.5\columnwidth]{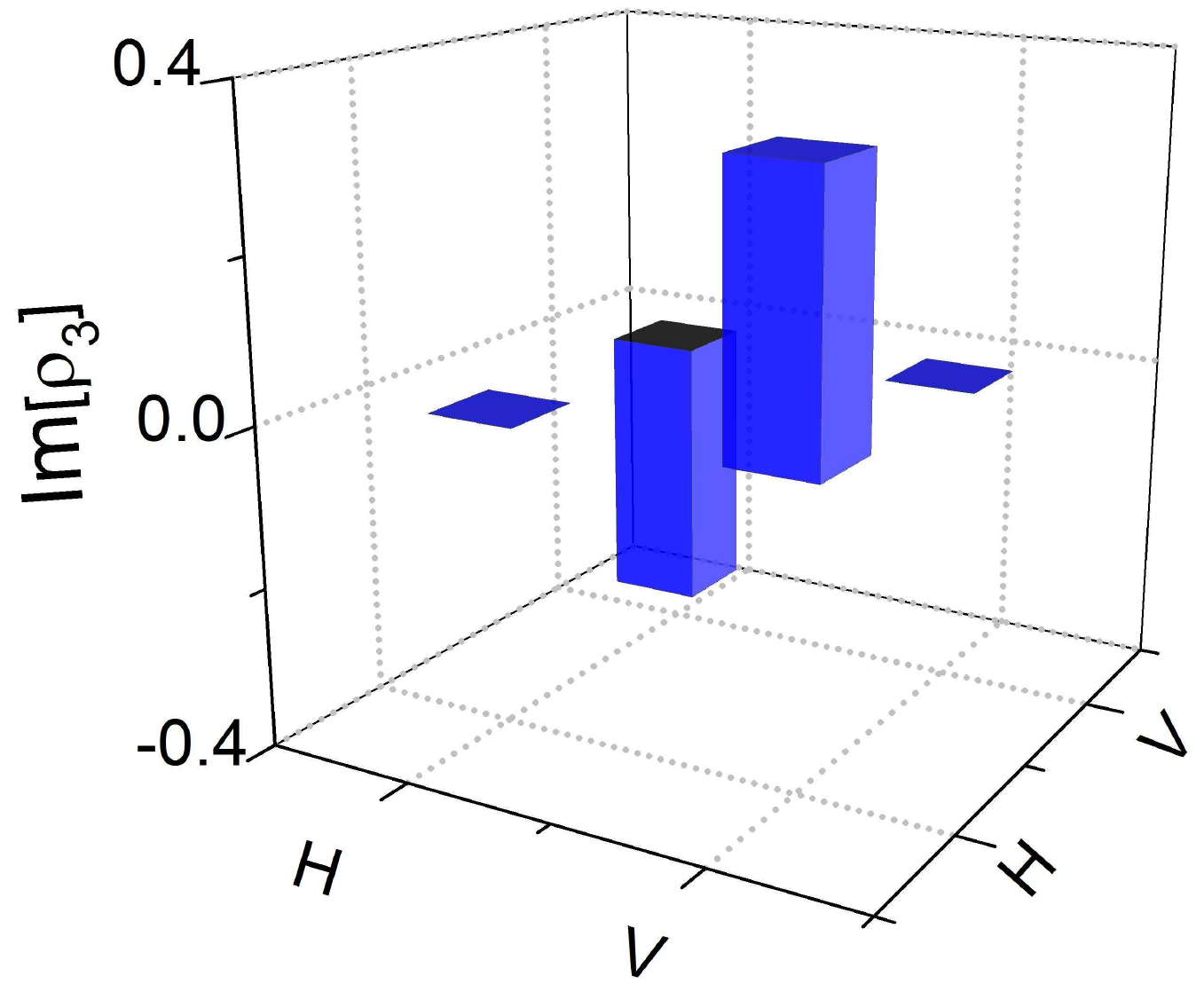}
\includegraphics[width=0.5\columnwidth]{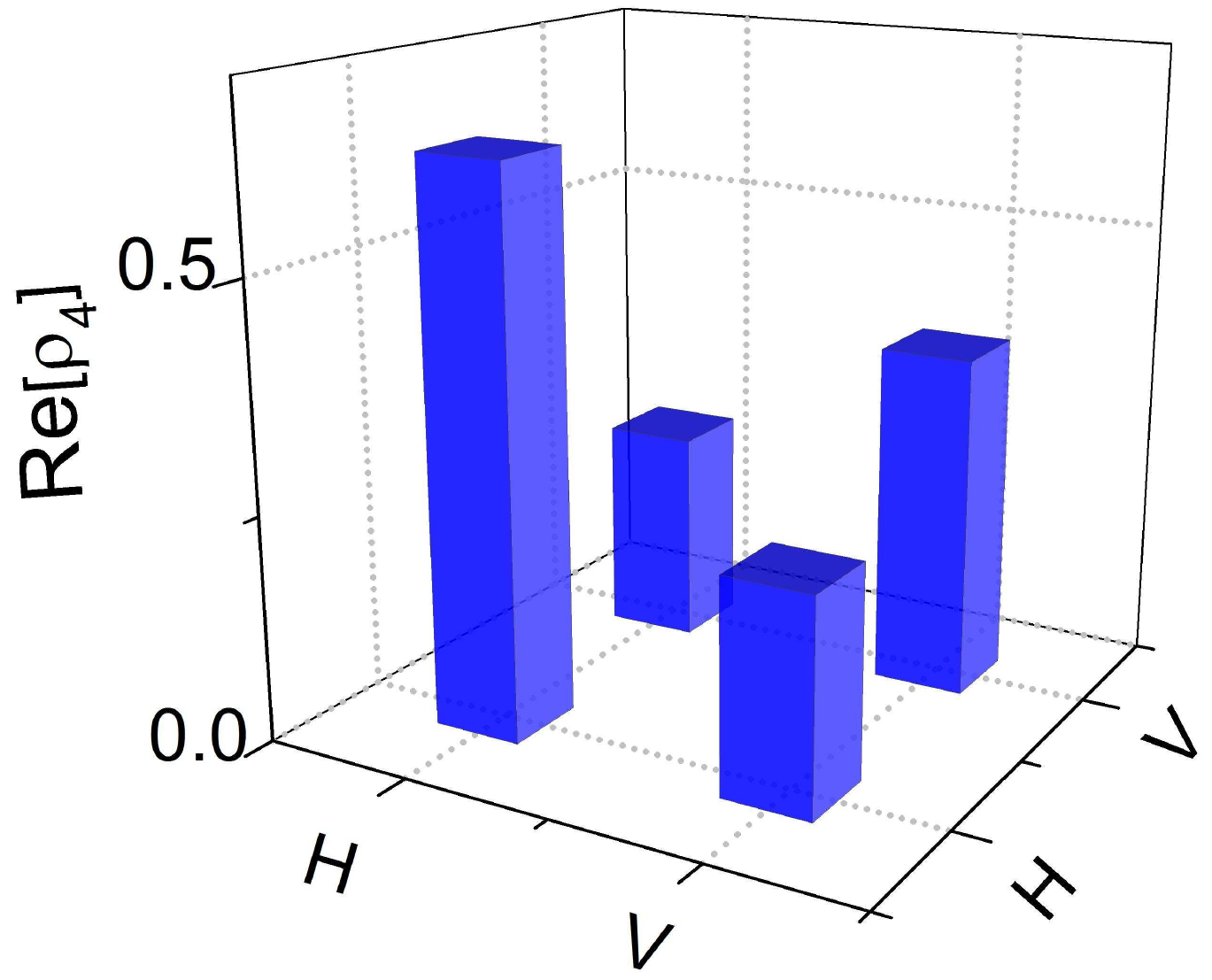}
\includegraphics[width=0.5\columnwidth]{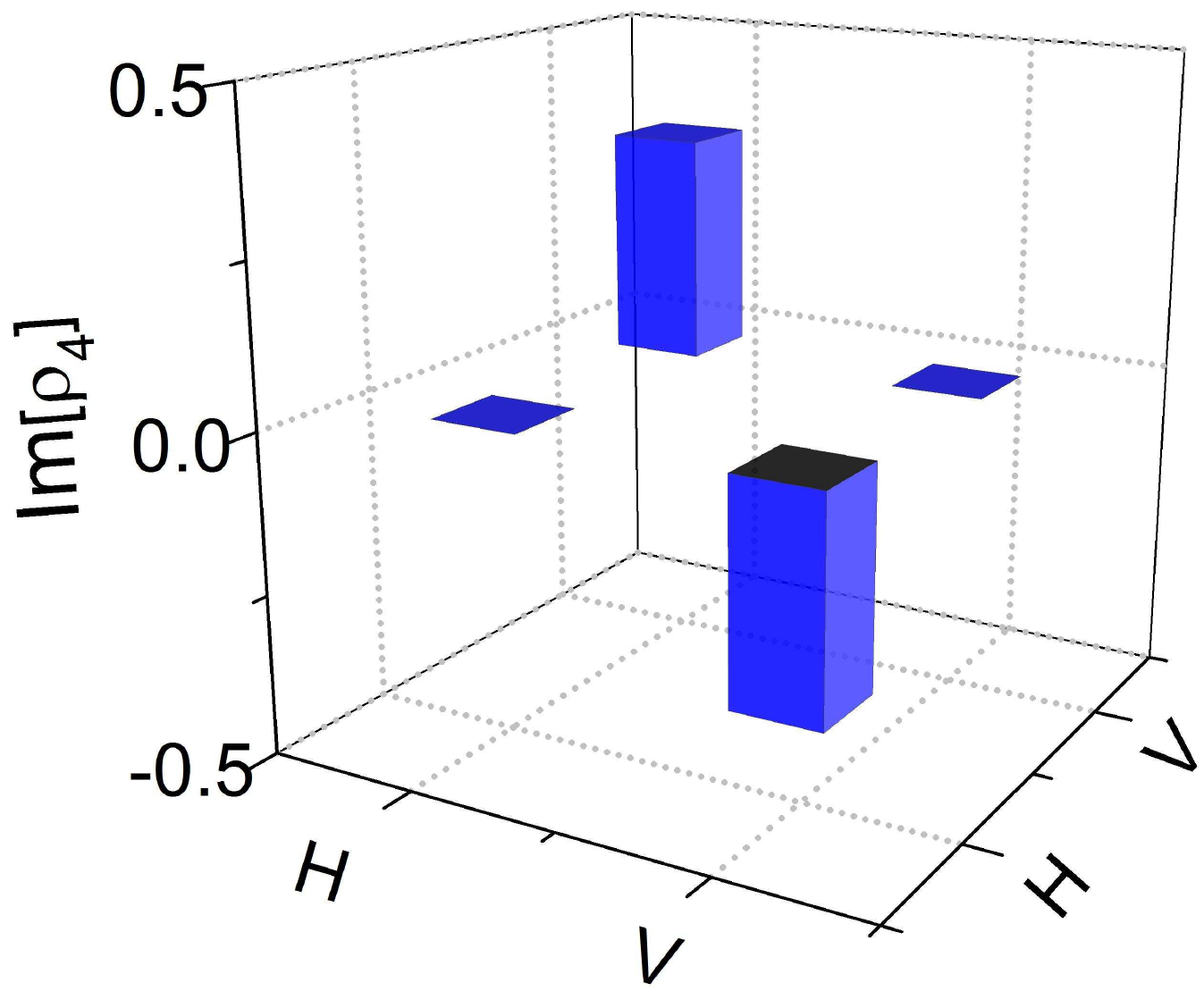}
\caption{(Color online.) Real and imaginary parts of the density matrices $\rho_i$ measured for four different initial
coin states $\ket{\psi^i}$ with $i=1,2,3,4$.}\label{supp}
\end{figure*}

This single qubit SIC-POVM can be realized by using a six-step photonic quantum walk. The initial coin state can be prepared by a HWP or quatre-wave plate (QWP) right after the PBS. For example, the initial coin states $\ket{\psi^1}$ and $\ket{\psi^2}$ can be prepared by a HWP with angles set to $\theta^H=45^o$ and $-17.63^o$, and the other two initial coin states $\ket{\psi^3}$ and $\ket{\psi^4}$ can be prepared by setting angle of QWP to $\theta^Q=-152.63^o$ and $117.37^o$, respectively. The single-qubit rotation realized by QWP is
\begin{equation}
R_\text{QWP}(\theta^Q)=\begin{pmatrix}\cos^2\theta^Q+\text{i} \sin^2 \theta^Q & (1-\text{i})\sin\theta^{Q}\cos\theta^Q\\
(1-\text{i})\sin\theta^Q\cos\theta^Q & \sin^2\theta^Q+\text{i}\cos^2\theta^Q \end{pmatrix},
\end{equation}
where $\theta^Q$ is the angle between the optic axes of QWP and horizontal direction.

The site-dependent coin rotations can be realized by HWP and QWP with specific angles show in Table II placed in certain modes.

\begin{table}[htbp]
\begin{tabular}{l|l|l|l|l|l|l}
  \hline
   & $\theta^H_0$ & $\theta^Q_0$ & $\theta^H_1$ & $\theta^Q_1$& $\theta^H_{-1}$ & $\theta^Q_{-1}$\\ \hline \hline
  step 1 & $-$ & $-$ & $-$& $-$ & $-$ & $-$ \\ \hline
  step 2 & $-$ & $-$ & $-22.5^o$ & $-$ & $45^o$ & $-$ \\ \hline
  step 3 & $67.5^o$ & $-$ & $-$ & $-$ & $-$ & $-$ \\
  \hline
  step 4 & $-$ & $-$ & $17.63^o$& $-$ & $45^o$ & $-$ \\ \hline
  step 5 & $52.5^o$ & $45^o$ & $-$ & $-$ & $-$ & $-$ \\ \hline
  step 6 & $-$ & $-$ & $-$ & $-$ & $45^o$ & $-$ \\
  \hline
\end{tabular}
\caption{The setting parameters of the HWPs and QWPs which are used to realize site-dependent coin rotations for the six-step quantum walk. The subscripts denote the optical modes where the wave plates are placed and ``$-$" means no corresponding wave plate is used.}
\end{table}

The measured probability distribution of the walker $P(x)$ and
1-norm distance $d$ from the theoretical predictions are shown corresponding to four different initial coin states are shown in Table III.

\begin{table}[htbp]
\begin{tabular}{l|l|l|l|l|l}
  \hline
  $ $ & $P(0)$ & $P(2)$& $P(4)$ & $P(6)$ & $d$ \\
  \hline\hline
  $\ket{\psi^1}$ & $0.3246(37) $ & $0.3277(38)$ &  $0.3327(38)$ & $0.0149(07)$ & $0.0149(33)$ \\ \hline
  $\ket{\psi^2}$ & $0.3398(38)$ & $0.3135(36)$  & $0.0345(11)$ & $0.3123(36)$ & $0.0401(32)$ \\ \hline
  $\ket{\psi^3}$ & $0.0335(10)$ & $0.3137(36)$ & $0.3432(38)$ & $0.3104(36)$ & $0.0425(32)$ \\ \hline
  $\ket{\psi^4}$ & $0.3158(36)$ & $0.0329(10)$ & $0.3419(38)$  & $0.3094(35)$ & $0.0415(32)$ \\ \hline
\end{tabular}
\caption{The measured probability distribution of the walker $P(x)$ and
1-norm distance $d$ from the theoretical predictions are shown corresponding to four different
initial coin states. Error bars indicate the statistical uncertainty.}
\end{table}

Moreover given a SIC-POVM
$\{\ket{\xi^i}\bra{\xi^i}/2\}$ ($i=1,2,3,4$) with the measured probability distribution the initial coin state can be reconstructed. The real and imaginary parts of the density matrices $\rho_i=\rho_i^{Re}+\text{i}\rho^{Im}_i$ corresponding to the states $\ket{\psi^i}$ are shown as
\begin{align}
&\rho_1^{Re}=\begin{pmatrix}
           0.0299 & 0.0093 \\
           0.0093 & 0.9701 \\
        \end{pmatrix},
        \rho_1^{Im}=\begin{pmatrix}
           0 & 0.0038 \\
           -0.0038 & 0 \\
        \end{pmatrix}\nonumber\\
&\rho_2^{Re}=\begin{pmatrix}
           0.6247 & -0.4132 \\
           -0.4132 & 0.3753 \\
        \end{pmatrix},
        \rho_2^{Im}=\begin{pmatrix}
           0 & -0.0322 \\
           0.0322 & 0 \\
        \end{pmatrix}\nonumber\\
&\rho_3^{Re}=\begin{pmatrix}
           0.6209 & 0.2386 \\
           0.2386 & 0.3791 \\
        \end{pmatrix},
        \rho_3^{Im}=\begin{pmatrix}
           0 & 0.3432 \\
           -0.3432 & 0 \\
        \end{pmatrix}\nonumber\\
&\rho_4^{Re}=\begin{pmatrix}
           0.6188 & 0.2370 \\
           0.2370 & 0.3812 \\
        \end{pmatrix},
        \rho_4^{Im}=\begin{pmatrix}
           0 & -0.3465 \\
           0.3465 & 0 \\
        \end{pmatrix}
\end{align}
Figure~\ref{supp} shows the histograms of the real and imaginary parts of
the density matrices obtained. The fidelities of the measured density matrices with respect to the theoretically predicted states (i.e., initial coin states) are $0.9701\pm0.0108$, $0.9311\pm0.0131$, $0.9330\pm0.0086$, and $0.9342\pm0.0102$, respectively.

\end{document}